\documentstyle [12pt] {article}

\def\L{{\cal L}}

\begin{document}
\baselineskip=24pt

\title{{\normalsize {\bf SELF-SIMILAR STATIC SOLUTIONS ADMITTING A TWO-SPACE
OF CONSTANT CURVATURE}}}

\author{ {\normalsize J. Carot, and A.M. Sintes }}
\date{} 
\maketitle
\centerline{ Departament de F\'{\i}sica}
\centerline{ Universitat de les Illes Balears}
\centerline{ E-07071 Palma de Mallorca. SPAIN}

\vskip4cm

\centerline{ \Large{ \bf Abstract}}
\vskip1cm
A recent result by Haggag and Hajj-Boutros \cite{Haggag} is reviewed within
the framework of self-similar space-times, extending, in some sense, their
results and presenting a family of metrics consisting of all the static
spherically symmetric perfect fluid solutions admitting a homothety.

 \vskip1cm


\setcounter{page}{0}
\newpage
In a recent paper,  Haggag and Hajj-Boutros \cite{Haggag} presented a static,
spherically symmetric perfect fluid solution with a stiff-matter type
equation of state (i.e.: $p=\mu$). By means of a few clever changes of
coordinates, the authors reduce the problem to that of solving a non-linear,
second order differential equation, whose polynomic solutions they
investigate showing that only three such solutions exist, two of them being
vacuum (flat Minkowski space-time and Schwarzschild solution) and the third
one being that leading to the new metric referred to above, henceforth called
HHB solution.

The purpose of this  letter is to give all the static, spherically symmetric
perfect fluid solutions admitting a homothety. This family can be completely
characterized by means of a real parameter $\gamma$ (arising quite naturally
from the equation of state for these fluids, see below), which must be in the
interval $[1,2]$ in order to satisfy energy conditions. The two limiting
values of $\gamma$, namely $\gamma=1$ and $\gamma=2$ correspond to Minkowski
flat space-time and to the HHB solution respectively.

A few remarks concerning the similarity group and its action are in order
here. It is a well known fact that an $r$-parameter group of homotheties $H_r$
(in which at least one proper homothety exists) always admits an
$(r-1)$-parameter subgroup of isometries $G_{r-1}$. Now, the maximal
dimension of the group   of homotheties that a perfect fluid space-time may
admit is $r=7$, in which case it is one of the special Robertson-Walker
space-times \cite{Hall90}, and therefore they are all known. The case $r=6$
is not compatible with an energy-momentum tensor  of the perfect fluid
type; thus, apart from the special Robertson-Walker solutions mentioned above,
the highest dimension of the group of homotheties that a perfect fluid
space-time may admit is $r=5$. In such case, the associated isometry subgroup
$G_4$ has necessarily 3-dimensional non-null orbits \cite{Hall90}. Notice that
this is precisely the case we are interested in. We shall not treat here the 
case in full generality, namely; studying all perfect fluid space-times
admitting an $H_5$ of homotheties, since this would be beyond the purpose of
this letter, but we shall restrict ourselves to the case when the subgroup
$G_4$ has timelike orbits $T_3$ and the subgroup $G_3$ that it necessarily
contains \cite{Kramer} has two-dimensional orbits. Everything else follows
from these assumptions and the field equations. For further information on
groups of homotheties and related issues, we refer   the reader to 
\cite{Hall88a,Hall88} and \cite{Hall90b}.
 \hfill\break

We start with a space-time that contains a non-null two-space of constant
curvature (i.e.: there exists a three-parameter isometry group $G_3$ acting on
this two-space). In this case the orbits $V_2$ admit orthogonal surfaces in $M$
\cite{Schmidt}. By performing a coordinate transformation in the two-spaces
orthogonal to the Killing orbits the space-time metric can be put into
diagonal form:
\begin{equation}
ds^2= A^2(r,t)(-dt^2 + dr^2) + B^2(r,t)(d\theta^2+f^2(\theta,k)d\phi^2)
\end{equation}
\begin{equation}
f(\theta,k)= \left\{ \begin{array}{ll}
\sin \theta & k =+1 \\
\theta & k =0 \\
\sinh \theta & k =-1. 
\end{array}
\right.
\end{equation}
where we have restricted ourselves to the case of spacelike Killing
orbits, since perfect fluid and dust solutions cannot admit a group $G_3$
on two-dimensional timelike orbits \cite{Kramer}.

Using the Jacobi identities and the fact that the Lie bracket of a proper
homothetic vector field (HVF) and a Killing vector (KV) is a KV it can be
easily shown that the HVF $X$ must be of either one of the  following forms:
\begin{eqnarray}
(I) \quad X & = & X^t(r,t)\partial_t+ X^r(r,t)\partial_r ,\quad
k=-1,0,1,\label{s2}\\
(II) \quad X & = &X^t(r,t)\partial_t+ X^r(r,t)\partial_r
- \theta\partial_{\theta}, \quad k=0. \label{s3}
\end{eqnarray}
Now by using isotropic coordinates, one finds that static metrics can be
expressed as \cite{Kramer}:
\begin{equation}
ds^2= -A^2(r)dt^2 +  B^2(r)[dr^2
+r^2(d\theta^2+f^2(\theta,k)d\phi^2)], \label{s4}
 \end{equation}
where $\partial_t$ is the hypersurface orthogonal timelike KV.

In this coordinate chart, the HVF in (\ref{s2}) and (\ref{s3}) takes the
following forms:
\begin{eqnarray}
(I) \quad X & = & nt\partial_t+ R(r)\partial_r ,\quad
k=-1,0,1,\label{s5}\\
(II) \quad X & = & nt\partial_t+ R(r)\partial_r
- \theta\partial_{\theta}, \quad k=0, \label{s6}
\end{eqnarray}
where $n$ is a constant.

The homothetic equation $\L_X g_{ab}=2g_{ab}$ specified to the components
$rr$ and $\theta\theta$ of the metric (\ref{s4}), gives:
\begin{equation}
R_{,r}-{R\over r} - {X^{\theta}}_{,\theta}=0,
\end{equation}
and integrating, one gets:
\begin{eqnarray}
(I) \quad X & = & nt\partial_t+ qr\partial_r ,\quad
k=-1,0,1,\label{s8}\\
(II) \quad X & = & nt\partial_t+ (-r\ln r +cr)\partial_r
- \theta\partial_{\theta}, \quad k=0, \label{s9}
\end{eqnarray}
where $c$  and $q$ ($\not= 0$) are constants. \hfill\break

{\bf Case ($I$)}

By means of the coordinate transformation $\, \hat r= r^{1/q}$, the HVF and
the metric can be written as
\begin{equation}
X = nt\partial_t+ \hat r\partial_{\hat r} ,
\end{equation}
\begin{equation}
ds^2= -\hat A^2(\hat r)dt^2 + \hat B^2(\hat r)[q^2 d\hat r^2
+\hat r^2(d\theta^2+f^2(\theta,k)d\phi^2)].
 \end{equation}
The metric functions can be determined via the homothetic equations, that
gives:
\begin{equation}
\hat B={\rm constant},\qquad \hat A \propto \hat r^{1-n}.
\end{equation}
Defining a new radial coordinate $r$ as $r=\hat r \hat B$, one can come to
the following simple forms for $X$ and the metric:
\begin{equation}
X = nt\partial_t+ r\partial_r ,
\end{equation}
\begin{equation}
ds^2= -r^{2(1-n)}dt^2 + q^2 dr^2
+ r^2(d\theta^2+f^2(\theta,k)d\phi^2). \label{s12}
 \end{equation}
\hfill\break

{\bf Case ($II$)}

Imposition of the homothetic equations specified to the metric (\ref{s4})
and to the HVF (\ref{s9}), leads directly to:
\begin{equation}
ds^2= -(-\ln r +c)^{2(n-1)}dt^2 +{b^2\over r^2 (-\ln r +c)^4}\left[ dr^2
+ r^2(d\theta^2+\theta^2d\phi^2)\right], \label{s13}
 \end{equation}
$b$ and $c$ being constants.\hfill\break

From the expressions (\ref{s12}) and (\ref{s13}) of the metric, it is
immediate to see that the components $tt$ of their respective Einstein
tensors are negative for $k=-1$ and $0$ (i.e.: hyperbolic and flat
two-spaces) and therefore cannot verify energy conditions. Thus, it only
remains to study  the spherically symmetric case.

In this latter case, the field equations for a perfect fluid matter content
lead to the metric:
\begin{equation}
ds^2= -r^{4-4/\gamma}\,dt^2 + \left({\gamma^2 +4\gamma-4\over \gamma^2}\right)
dr^2 + r^2(d\theta^2+\sin^2\theta d\phi^2), \label{s14}
 \end{equation}
These metrics were already found, following a completely different approach,
by Iba\~nez et al.\cite{Ibanez}, and particular cases of them can be also
found in Misner et al.\cite{Misner} (which are particular cases of Tolman class
$VI$ solutions). Some particular cases (when the HVF is orthogonal to the
fluid flow) were also studied by Herrera   et al.\cite{Herrera84}.

The matter variables being
\begin{equation}
\mu={1\over r^2}\left({4\gamma-4 \over \gamma^2 +4\gamma-4}\right),
\end{equation}
\begin{equation}
p=(\gamma-1)\mu
\end{equation}
as one would have expected from $p$ and $\mu$ being functions of $r$ alone
(and therefore, by the implicit function theorem, the fluid has a
barotropic equation of state) and the space-time being self-similar
\cite{Wainwright}. The HVF takes then the form:
\begin{equation}
X = {2-\gamma \over \gamma}t\partial_t+ r\partial_r .
\end{equation}
These are all the static, spherically-symmetric self-similar perfect fluid
solutions. They are shear-free and have null volume expansion since the
four-velocity $u$ of the fluid is parallel to the timelike KV, the
vorticity is also zero  (since $u$ is orthogonal to the orbits $S_2$ of the
$G_3$ they contain); and the fluid has non-geodesic flow.

The particular case, $\gamma=2$, is the HHG solution \cite{Haggag} and in
this case the HVF $X$ becomes orthogonal to the fluid four-velocity, and for
$\gamma=1$ the space-time is obviously flat.

It is interesting to notice that static and self-similar solutions
admitting  a two-space of constant curvature can only be spherically
symmetric (irrespectively of the matter content) and that an $H_5$ static
space-time with a $G_3$ acting on spacelike orbits, necessarily contains an
$H_4$.


\newpage

\begin{thebibliography}{99}
\bibitem{Haggag} S. Haggag and J. Hajj-Boutros, Class. Quantum. Grav.,  {\bf
11}(1994)L69.
\bibitem{Hall90} G.S. Hall and J.D. Steele, Gen. Rel. Grav., {\bf
22}(1990)457.  
\bibitem{Kramer} D. Kramer, H. Stephani, M.A.H. MacCallum and E. Herlt. {\em
Exact Solutions of Einstein's Field Equations}. Deutscher Verlag der
Wissenschaften, Berlin (1980).
\bibitem{Hall88a} G.S. Hall, Class. Quantum. Grav.,  {\bf
5}(1988)L77.
\bibitem{Hall88} G.S. Hall, Gen. Rel. Grav., {\bf 20}(1988)671.
\bibitem{Hall90b} G.S. Hall, J. Math. Phys. {\bf 31}(1990)1198.
\bibitem{Schmidt} B.G. Schmidt, Z. Naturforsch.  {\bf 22a}(1967)1351.
\bibitem{Wainwright} J. Wainwright, {\it Self-similar solutions of
Einstein's equations} in {\em Galaxies, Axisymmetric systems and
Relativity}, ed. M.A.H. MacCallum, Cambridge (1985) C.U.P.
\bibitem{Ibanez} J. Iba\~nez and J.L. Sanz, 
J. Maths. Phys., {\bf 23}(1982)1364.
\bibitem{Misner} C.W. Misner and H.S. Zapolsky, Phys. Rev. Lett., {\bf
12}(1964)635.
\bibitem{Herrera84} L. Herrera, J. Jimenez, L. Leal, J. Ponce de Leon, M.
Esculpi and V. Galina, J. Maths. Phys., {\bf 25}(1984)3277.
 \end{thebibliography}
 \end{document}